\documentclass[%
 preprint,
 superscriptaddress,
 amsmath,amssymb,
 aps,
floatfix,
]{revtex4-2}

\usepackage[english]{babel}

\usepackage[letterpaper,top=2cm,bottom=2cm,left=3cm,right=3cm,marginparwidth=1.75cm]{geometry}

\usepackage{amsmath}
\usepackage{graphicx}
\usepackage[colorlinks=true, allcolors=blue]{hyperref}
\usepackage{xcolor}

\begin{document}
\title{Direct measurement of a $\sin(2\varphi)$ current phase relation in a graphene superconducting quantum interference device}

\author{Simon Messelot}
\author{Nicolas Aparicio}
\author{Elie de Seze}
\author{Eric Eyraud}
\author{Johann Coraux}
\affiliation{Univ. Grenoble Alpes, CNRS, Grenoble INP, Institut N\'eel, 38000 Grenoble, France}
\author{Kenji Watanabe}
\affiliation{Research Center for Electronic and Optical Materials, National Institute for Materials Science, 1-1 Namiki, Tsukuba 305-0044, Japan}
\author{Takashi Taniguchi}
\affiliation{Research Center for Materials Nanoarchitectonics, National Institute for Materials Science,  1-1 Namiki, Tsukuba 305-0044, Japan}
\author{Julien Renard}\email{julien.renard@neel.cnrs.fr}
\affiliation{Univ. Grenoble Alpes, CNRS, Grenoble INP, Institut N\'eel, 38000 Grenoble, France}


\begin{abstract}
In a Josephson junction, the current phase relation relates the phase variation of the superconducting order parameter, $\varphi$, between the two superconducting leads connected through a weak link, to the dissipationless current . This relation is the fingerprint of the junction. It is usually dominated by a $\sin(\varphi)$ harmonic, however its precise knowledge is necessary to design superconducting quantum circuits with tailored properties. Here, we directly measure the current phase relation of a superconducting quantum interference device made with gate-tunable graphene Josephson junctions and we show that it can behave as a $\sin(2\varphi)$ Josephson element, free of the traditionally dominant $\sin(\varphi)$ harmonic. Such element will be instrumental for the development of superconducting quantum bits protected from decoherence. 
\end{abstract}

\maketitle

\section*{Main text}

In superconducting circuits, nonlinearity such as the one provided by a Josephson junction is used as a resource for storing, writing and processing quantum information. This celebrated building block is indeed characterized by a nonlinear current phase relation (CPR) relating the supercurrent $I$ to the phase across the junction $\varphi$. Standard superconducting circuits use tunnel junctions for which the Josephson relation $I(\varphi) = I_C \sin(\varphi)$, where $I_C$ is the junction critical current, allows to predict with high accuracy the behavior of non-linear circuits such as Josephson parametric amplifiers or Qubits \cite{aumentado2020superconducting,krantz2019quantum}.  In recent years, there has been a rising interest for junctions made without a tunnel barrier, for instance using a gate-tunable semiconductor weak link \cite{larsen2015semiconductor,de2015realization,casparis2016gatemon,casparis2018superconducting,wang2019coherent,strickland2023characterizing,huo2023gatemon,hertel2022gate,zhuo2023hole,butseraen2022gate,sarkar2022quantum,phan2023gate}. In such junctions, with potentially highly transmitting Andreev channels, the CPR is more complex and includes higher order $\sin(2\varphi)$, $\sin(3\varphi)$, etc. harmonics \cite{golubov2004current}. Such higher order contributions, that are traditionally neglected in canonical tunnel junctions, were recently found to influence the spectrum of excited states of Qubits made with aluminum/aluminum oxide junctions \cite{willsch2024observation}. 
In the future, the precise knowledge of the harmonic content of the CPR could be used to design protected qubits \cite{larsen2020parity} or more generally highly tunable dissipationless nonlinearities \cite{schrade2023dissipationless}.

Various kinds of Josephson junctions (weak links) have been considered, for instance with a metal or a semiconductor. In these SNS junctions, higher order harmonics have been reported in the CPR with InAs nanowires \cite{spanton2017current}, InAs quantum wells \cite{zhang2024large, nichele2020relating,ciaccia2023gate}, bismuth nanowires \cite{murani2017ballistic,bernard2021evidence}, BiSbTeSe$_2$ topological insulator \cite{kayyalha2019anomalous, kayyalha2020highly}, ferromagnetic Bi$_{1-x}$ Sb$_x$ and other metal based $\pi$ junctions \cite{li2019zeeman,baselmans2002direct}, WTe$_2$ \cite{endres2023current}, as well as graphene \cite{chialvo2010current,nanda2017current}. There, the determination of the CPR relied on the conventional direct current (DC) bias method \cite{della2007measurement}, whereby the Josephson junction under investigation is connected in parallel to a second reference junction.

This method faces two main limitations: the higher-order harmonics can be hard to detect when the $\sin(\varphi)$ term is dominating. Also, to retrieve the CPR one usually needs to assume a fixed phase $\varphi_\mathrm{ref}$ in the reference junction, which is often questionable in real-life experimental systems \cite{endres2023current}, and disregarding $\varphi_\mathrm{ref}$ changes in the analysis can generate spurious higher-order harmonics in the CPR \cite{babich2023limitations}. Other methods relying on radiofrequency techniques have been developed, such as the detailed exploration of Shapiro steps \cite{ueda2020evidence,leblanc2023nonreciprocal,valentini2024parity,zhao2023time}, probing the microwave susceptibility \cite{murani2019microwave} or photon emission \cite{ciaccia2024charge}. These remain however mostly indirect methods to measure the CPR, not free of artefacts, which altogether can make quantitative measurements challenging.\\

In this letter, we demonstrate a method to control and read simultaneously the CPR of a Josephson element made with a superconducting quantum interference device (SQUID) based on graphene Josephson junctions. We use a double SQUID structure. The first, symmetric SQUID constitutes the tunable Josephson element under study. Its CPR is controlled using a magnetic flux and gate voltages. An additional loop including a third Josephson junction serves as reference arm for conventional DC-biased CPR measurements. The value of the magnetic flux $\phi_1$ enclosed within the symmetric SQUID controls the amplitudes of the various harmonics. In the frustrated regime, \textit{i.e.} $\phi_1 = \frac{\phi_0}{2}$, where $\phi_0$ is the flux quantum, this circuit naturally selects the even-order harmonics of the Josephson element, isolated from the otherwise dominant first harmonic. This allows direct visualisation of $\sin(2\varphi)$ oscillations of the CPR and quantitative measurement of this second order harmonic amplitude. Like with other methods, our measurement scheme relies on a reference branch. We present a procedure that quantifies and compensates for the deviation from the standard assumption that the phase of the reference junction is fixed. We use this method to determine quantitatively the relevant circuit parameters from experimental data.

\begin{figure*}
\centering
\includegraphics[width=1\linewidth]{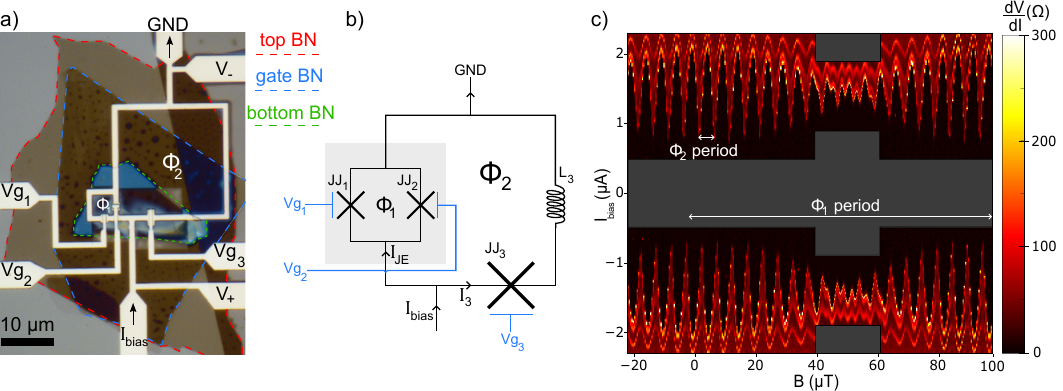}
\caption{\label{fig:Fig1} \textbf{a)} Optical micrograph of the double SQUID device showing the 2D materials heterostructure and the superconducting aluminum circuit. The graphene Josephson junctions are located under the top gates. \textbf{b)} Equivalent electrical circuit of the device, including the Josephson element whose CPR is measured (gray area), the reference branch (right), and the top gates (blue).  \textbf{c)} Differential resistance map versus bias current and magnetic field when all gates are at 0 V. The grey area indicate data points that were not measured.}
\end{figure*}

Our double SQUID device (\textcolor{blue}{Fig.~\ref{fig:Fig1}.a}) includes three graphene Josephson junctions of dimensions $1 \times 0.4$ µm$^2$ (JJ$_1$ and JJ$_2$) and $2.5 \times 0.4$ µm$^2$ (JJ$_3$), two $4.5 \times  4.5$ µm$^2$ and $20.5 \times 20.5$ µm$^2$ superconducting loops with 1~µm wire width, three top gate electrodes, and electrical contacts for DC-biased measurements. 

The Josephson junctions are made with monolayer graphene encapsulated within two hexagonal boron-nitride (hBN) layers, and are contacted at their edges \cite{wang2013one} by 5/60 nm Ti-Al superconducting electrodes. The critical current of the junctions can be tuned using gate voltages applied to top gate electrodes, insulated from the junctions using an additional hBN layer. Details about the fabrication methods and JJ structure are provided in the Supplemental Material \cite{SM}.

The device is cooled down to 30 mK in a dilution refrigerator. We perform bias current sweeps while measuring  the voltage drop $V$ across the device in a 4-probe configuration with two additional $V_+$ and $V_-$ electrodes (see \textcolor{blue}{Fig.~\ref{fig:Fig1}.a}), a lock-in amplifier and a digital voltmeter. Magnetic flux is applied using a superconducting coil on top of the sample holder, yielding 8.7 µT$\cdot$mA$^{-1}$, as determined from the periodicity of the measurements versus the applied current (geometrical theoretical value is $\simeq$ 8 µT$\cdot$mA$^{-1}$).\\
The Al-based superconducting circuit represented in \textcolor{blue}{Fig.~\ref{fig:Fig1}.b} consists of two main parts. First, a small-area SQUID that includes JJ$_1$ and JJ$_2$, which intercepts a magnetic flux $\phi_1$. This is the tunable Josephson element (JE) of which we will control the CPR $I_{JE}(\varphi)$ with the flux $\phi_1$: 

\begin{equation}
    I_{JE}(\varphi,\phi_1) = I_1(\varphi+ 2\pi \frac{\phi_1}{\phi_0})   + I_2(\varphi )
\end{equation}

where $I_1(\varphi_1)$ and $I_2(\varphi_2)$ are the CPRs of JJ$_1$ and JJ$_2$ whose critical currents $I_{c1}$ and $I_{c2}$ are controlled by gate voltages $V_{g1}$ and $V_{g2}$. The Josephson element is connected to a larger loop, controlled by a flux $\phi_2$, which contains junction JJ$_3$ of large critical current $I_{c3}$, tunable using $V_{g3}$. Since we use a single coil to produce the magnetic field, $\phi_1$ and $\phi_2$ vary together with a linear relationship determined by the surface area ratio of the two loops. In the large-area SQUID the phase $\varphi$ and the phase of the reference junction $\varphi_{ref}$ are linked by:

\begin{equation}
    \varphi =  \varphi_{ref} + 2\pi \frac{\phi_2}{\phi_0}
    \label{eqn:phaseRel}
\end{equation}

This way, the reference branch enables the measurement of the CPR of the Josephson element by varying $\phi_2$ and measuring the critical current of the device. Provided that $I_{c3}$ is much larger than the Josephson element's critical current, we can assume that $\varphi_{ref}$ is fixed, at the critical current of full device, to $\varphi_{ref} = \varphi_{max}$ with $\varphi_{max}>\frac{\pi}{2}$ because of the use of a SNS reference junction, with a forward skewed CPR \cite{golubov2004current}. We can then apply the conventional DC-biased CPR measurement method \cite{della2007measurement} to the tunable Josephson element. 

For completeness, we also consider the contribution of the inductance of the reference branch ($L_3$).  In the Supplemental Material \cite{SM} we show that its effect is small and reduces to a bias-current dependent shift of $\phi_2$. Other inductances have even smaller effects due to both shorter length and smaller circulating currents, and can be safely neglected. Finally, we estimated the shunt capacitance of the device, including the dominant contribution of the top gate electrodes, to be of the order of 25 fF. The junction's quality factor $Q$ has then a value between 0.6 and 0.8 in our measurements depending on the gate voltages and the device is thus in the slightly overdamped regime \cite{tang2022overdamped}, confirmed by the absence of hysteretic behavior. In the following, we will thus identify the critical current to the experimentally measured switching current. \\

A typical critical current measurement is presented in \textcolor{blue}{Fig.~\ref{fig:Fig1}.c)}, for which top gates were all kept grounded. The device's critical current is measured using sweeps of the bias current (from low to high values) at varying magnetic field values. In \textcolor{blue}{Fig.~\ref{fig:Fig1}.c)}, the superconducting regions ( $\frac{dV}{dI} = 0$, black in the figure) are delimited by the critical current (peaks in $\frac{dV}{dI} = 0$, bright in the figure). Starting from zero magnetic flux, we observe fast variations of the critical current with a 5.1 µT period. Each period corresponds to an additional flux $\Delta \phi_2 = \phi_0$ in the large loop (hence $\Delta\varphi=2\pi$). In other words they are an image of the CPR of the Josephson element, $I_{\mathrm{JE}}(\varphi)$. These fast variations are slowly modulated within an envelope of period 102 µT that stems from the evolution of the Josephson element CPR with $\phi_1$. As $\phi_1$ approaches the frustration point $\phi_1 =\frac{\phi_0}{2}$, the dominant first harmonic of the CPR is suppressed, making the critical current through the Josephson element smaller. Similar multi-period critical current oscillations were also observed in rhombi chains \cite{gladchenko2009superconducting}. The ratio of the slow and fast periods is $\phi_2 \simeq 20.0 \phi_1$, consistent with the ratio between the loops area.\\

The device enables then to probe the CPR of the Josephson element using $\phi_2$ for different values of $\phi_1$. $\phi_1$ and $\phi_2$ are codependent but the ratio between their periods is large, so that $\phi_1$ variations are small over one period of $\phi_2$. The ratio between the reference junction critical current $I_{c3}$ and the Josephson element critical current is however not large enough in \textcolor{blue}{Fig.~\ref{fig:Fig1}.c)}, typically about 10 close to the frustration point $\phi_1 =\frac{\phi_0}{2}$. The assumption that the reference junction is fixed does not fully hold and this plot does not then represent the true CPR of the Josephson element. Also, we expect JJ$_1$ and JJ$_2$ to be asymmetric without proper tuning of the critical currents $I_{c1}$ and $I_{c2}$, making a full suppression of the $\sin(\varphi)$ term impossible. In the following, we take advantage of the critical current tunability for junctions JJ$_1$ and JJ$_2$ to achieve two necessary conditions to measure a pure $\sin(2\varphi)$ Josephson element CPR: moderate the total Josephson element critical current and ensure balanced $I_{c1}$ and $I_{c2}$.\\

\begin{figure*}[t]
\centering
\includegraphics[width=1\linewidth]{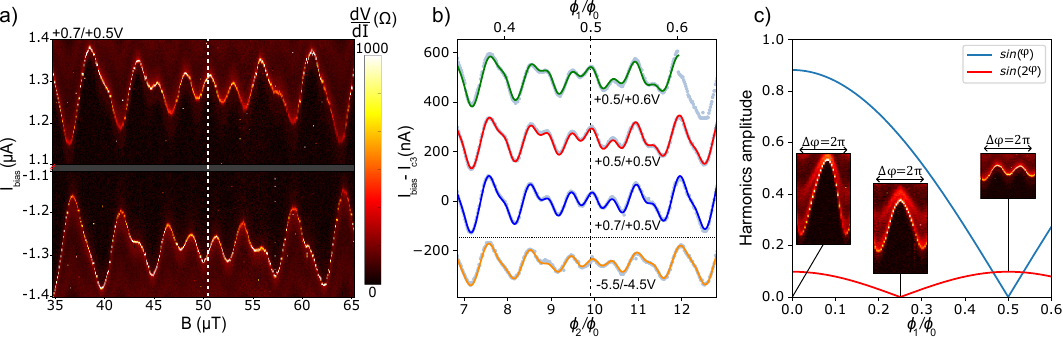}
\caption{\label{fig:Fig2} \textbf{a)} Differential resistance map around the frustration point (white vertical dashed line), for symmetric weakly n-doped JJs. \textbf{b)} Josephson element's critical current versus magnetic flux (top axis, $\phi_1$, small-loop portion, i.e. Josephson element's flux ; bottom axis, $\phi_2$, large-loop portion, i.e. CPR measurement flux). Experimental data (dot) and fits using Eq.~\ref{eq:analytic} (color solid lines) are shown. $V_{g1}$ and $V_{g2}$ gate voltages are indicated below the corresponding curve. Red, green and blue: weakly n-doped graphene. Yellow: Strongly p-doped graphene. \textbf{c)} Theoretical contributions of the harmonics of a Josephson element based on a graphene Josephson junctions SQUID, normalized to the zero-flux critical current (junctions with $T = 0.6$). Insets: CPR measured at specific flux values.}
\end{figure*}

We show in \textcolor{blue}{Fig.~\ref{fig:Fig2}.a)} a detailed measurement at the frustration point $\phi_1 =\frac{\phi_0}{2}$ at the symmetric sweet-spot, i.e. for $I_\mathrm{c1}=I_\mathrm{c2}$. To reach this configuration, we weakly doped the two graphene channels with electrons, with $V_\mathrm{g1}=+0.7$~V and $V_\mathrm{g2}=+0.5$~V above their charge neutrality points. The corresponding evolution of the Josephson element's critical current with $\phi_2$ is displayed in Fig. 2b. For these measurements we used a large gate voltage $V_{g3} = +6.5$ V to ensure a large critical current of the reference junction $I_{c3} = 1285$ nA. In these conditions, we observe that the periodicity of the critical current is reduced by a factor 2. This demonstrates full cancellation of the first harmonic in the Josephson element's CPR leaving a dominant $\sin(2\varphi)$ term.\\

Quantitative analysis requires a model for fitting the critical current data to account for $\phi_1$ and $\phi_2$ varying together. Assuming  a fixed phase ($\varphi_{ref} = \varphi_{max}$) and a current $I_{c3}$ in the reference junction JJ$_3$ at the critical current, we can express the positive and negative critical currents  $I_{c}^{\pm}$:

\begin{equation}
    I_{c}^{\pm} = \pm I_{c3} + I_2(\pm\varphi_{max} + 2\pi \frac{\phi_2}{\phi_0}) + I_1(\pm\varphi_{max} + 2\pi \frac{\phi_2}{\phi_0} + 2\pi \frac{\phi_1}{\phi_0})
    \label{eq:analytic}
\end{equation}

We chose as model for all junctions the expression of the CPR of an SNS junction in the short and ballistic regime \cite{golubov2004current}:

\begin{equation}
    I(\varphi) = I' \frac{\sin(\varphi)}{ \sqrt{1-T \sin^2 (\varphi/2)} }
    \label{eq:SNSCPR}
\end{equation}

with $T$ the channels transparency and $I'$ sets the scale of the critical current. In the following, we write the junction's CPR as a series of harmonics $i$ of amplitude $A_i$ as $I(\varphi) = \sum_i A_i \sin(i \varphi)$. Finally, we fit both the positive and negative critical current data  $I_{c}^+$ and  $I_{c}^-$ with the same set of parameters.

We show the detailed evolution of the SQUID CPR at $\phi_1 =\frac{\phi_0}{2}$  in  \textcolor{blue}{Fig.~\ref{fig:Fig2}.b)} for varying gate voltages $V_{g1}$ and $V_{g2}$. 
From top to bottom, the critical current symmetry between JJ$_1$ and JJ$_2$ is improved. For the green curve ($V_\mathrm{g1}/V_\mathrm{g2}=+0.5$V$/+0.6$V), JJ$_1$ and JJ$_2$ are poorly balanced and we observe that the signature of the $\sin(2\varphi)$ term is mostly masked by a dominating first harmonic. The red curve ($V_\mathrm{g1}/V_\mathrm{g2}=+0.5$V$/+0.5$V) is close to balanced, first and second  harmonics are of similar amplitudes. Finally, for the blue curve ($V_\mathrm{g1}/V_\mathrm{g2}=+0.7$V$/+0.5$V), the Josephson element SQUID is symmetric, the $sin(2\varphi)$ term is dominant and the first harmonic is strongly suppressed. The fit yields  $T=0.53 \pm 0.02$ (resp. $T=0.63 \pm 0.015$) or $A_2/I_c = 0.094 \pm 0.004$ (resp.  $A_2/I_c = 0.117 \pm 0.003$) for JJ$_1$ (resp. JJ$_2$). We hence estimate that in the Josephson element at $\phi_1 =\frac{\phi_0}{2}$, the $\sin(\varphi)$ harmonic term is reduced to $13\%$ ($\pm 9\%$) of the $\sin(2\varphi)$ harmonic term, i.e. that the second harmonic contribution is about ten times larger than the first harmonic.\\
 
We also perform a similar measurement in a situation where graphene is strongly hole-doped, with $V_{g1} = -5.5$ V and $V_{g2}=-4.5$ V relative to the neutrality point (bottom curve in Fig.~\ref{fig:Fig2}.b). Large negative gate voltages are necessary to reach the same range of critical currents for JJ$_1$ and JJ$_2$ \cite{calado2015ballistic} (see Supplemental Material for more details about the junctions' control using gate voltage \cite{SM}). We observe a slightly smaller second harmonic term which is a consequence of reduced transparencies of the channels. We obtain  $T=0.49 \pm 0.025$ (resp. $T=0.50 \pm 0.025$) or $A_2/I_c = 0.086 \pm 0.0045$ (resp.  $A_2/I_c = 0.088 \pm 0.0045$) for JJ$_1$ (resp. JJ$_2$).\\

\textcolor{blue}{Figure~\ref{fig:Fig2}.d)} summarizes the expected evolution of the first and second harmonics weights of the CPR with $T=0.6$ and it displays, at some representative fluxes, the measured data. At $\phi_1 = 0$, the entire harmonic content of the CPR is present and we observe a skewed sine profile similar to reports on graphene Josephson junctions \cite{nanda2017current}. At $\phi_1 =\frac{\phi_0}{4}$, the CPR is close to a pure sine function because the $\sin(2\varphi)$ contributions of the 2 JJs cancel out and harmonics higher than 2 are small ($A_3 / I_c \simeq 1.6\%$ for a single channel at $T=0.6$). At $\phi_1 =\frac{\phi_0}{2}$ when all odd order harmonics are suppressed, we observe an almost pure $sin(2\varphi)$ current phase relation.\\

A common limitation of the DC bias method we use is the mapping of the Josephson element's phase $\varphi$ onto magnetic flux $\phi_2$. This is usually done using Eq.~\ref{eqn:phaseRel}, assuming $\varphi_{ref} = \varphi_{max}$. It is possible to work when this assumption is not valid, provided the existence of a model for the CPR. In the following, we investigate errors related to the $\varphi_{ref} = \varphi_{max}$ assumption and present a method to quantitatively characterize the device even for low $I_{c3}/I_{c,JE}$ ratio, such as in \textcolor{blue}{Fig.~\ref{fig:Fig1}.c)}.\\

If we relax the assumption of fixed reference JJ$_3$ phase, we can still write the critical current of the full device as the maximum of the sum of all junctions currents:

\begin{equation}
    I_{c}^+(\phi_1,\phi_2) = \max_{\varphi_ {ref}} \left[ I_3(\varphi_{ref}) + I_{JE}(\varphi_{ref} + 2\pi \frac{\phi_2}{\phi_0},\phi_1)  \right]
    \label{eq:maxEq}
\end{equation}

\begin{figure*}[t]
\centering
\includegraphics[width=8.6cm]{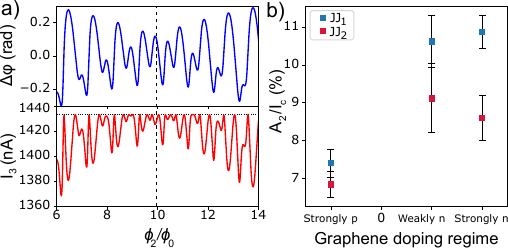}
\caption{\label{fig:Fig3}  \textbf{a)} Deviation of the reference phase $\varphi_{ref}$ from the maximum phase $\varphi_{max}$ (top) and deviation of the current $I_3$ from the critical current $I_{c3}$ (bottom). The values are computed using the data of \textcolor{blue}{Fig.~\ref{fig:Fig1}.c)} and the exact model of Eq.~\ref{eq:maxEq}. The dashed line shows $\phi_1 = \frac{\phi_0}{2}$.  \textbf{b)} Second CPR harmonic amplitude contribution to the critical current for different dopings, obtained from fit results using the complete model. In the analysis, contributions of JJ$_1$ and JJ$_2$ cannot be distinguished unambiguously.}
\end{figure*}

Using the model for the CPR given by Eq.~\ref{eq:SNSCPR}, we compute this expression numerically and fit our critical current data extracted from \textcolor{blue}{Fig.~\ref{fig:Fig1}.c)}, similarly to what was performed in Ref. \cite{jha2024large}. From the fit parameters, we derive $\Delta \varphi = \varphi_{ref} - \varphi_{max}$ and the current in the reference junction $I_3$ at the device critical current (\textcolor{blue}{Fig.~\ref{fig:Fig3}.a)}). These two plots represent, respectively, the X and Y errors in the CPR measurement when assuming a fixed reference phase. Around the frustration point, the deviations are significant with $\Delta \varphi$ as high as $\pm 0.17$ rad and $I_3$ dropping by up to 25 nA. It is possible to demonstrate that the phase (resp. current) errors can be expressed as: $\Delta \varphi \simeq - \frac{1}{I_{c,ref}} \frac{\partial I}{\partial \varphi}$ and $\Delta I \simeq - \frac{1}{2I_{c,ref}}  \left( \frac{\partial I}{\partial \varphi} \right)^2$  (see Supplemental Material \cite{SM}). The linear and quadratic dependence are apparent in {Fig.~\ref{fig:Fig3}.a)}. Using the same method, we evaluated that the deviations for the plots of \textcolor{blue}{Fig.~\ref{fig:Fig2}.b)} are limited to within $\pm 0.05$ rad and $2$ nA confirming the validity of the fixed-$\varphi_\mathrm{ref}$ assumption to derive the CPR in this case.\\

Finally, in \textcolor{blue}{Fig.~\ref{fig:Fig3}.b)} we summarize the $\sin(2\varphi)$ amplitudes we extracted from the different measurement at different doping levels, using the exact numerical method we presented. We find a $\sin(2\varphi)$ contribution to the CPR up to $10.9\%$ in the deeply n-doped regime, and a lower value in the deeply p-doped regime, about $7 \%$, due to lower interface transparencies. There is however no significant difference between weakly and strongly n-doped junctions, suggesting that the channel transparency most likely very quickly saturates as soon as the graphene is weakly n-doped. The complete analysis also confirms that the two Josephson junctions are not equivalent. This could have multiple causes related to the fabrication, including defects induced during the stacking of the different layers of the hBN/graphene/hBN heterostructure or inhomogeneous doping due to different top gate geometries.\\

In conclusion, the double SQUID device we used enables the control and readout of the CPR of a Josephson element,  taking advantage of the possibility to fine tune electrically the critical currents. This allowed us to directly show that a graphene SQUID can behave as a $\sin(2\varphi)$ Josephson element. Our method also facilitates the quantification of higher order harmonics contributions. In the future, decoupling between control flux $\phi_1$ and readout flux $\phi_2$ could be achieved using local flux lines. Our work paves the way to the future integration of such tunable $\sin(2\varphi)$ Josephson element in advanced qubit designs and the demonstration of protection from decay and dephasing.

\section*{A\lowercase{cknowledgments}}
This work was supported by the French National Research Agency (ANR) in the framework of the Graphmon project (ANR-19-CE47-0007). This work benefited from a French government grant managed by the ANR agency under the ‘France 2030 plan’, with reference ANR-22-PETQ-0003. K.W. and T.T. acknowledge support from the JSPS KAKENHI (Grant Numbers 21H05233 and 23H02052) and World Premier International Research Center Initiative (WPI), MEXT, Japan. We acknowledge the staff of the Nanofab cleanroom of Institut Néel for help with device fabrication. We acknowledge the work of J. Jarreau, L. Del-Rey and D. Dufeu for the design and fabrication of the sample holders and other mechanical pieces used in the cryogenic system. We thank E. Bonnet for help with the cryogenic system.  We thank A. Leblanc and F. Lefloch for discussions and comments.


%

\end{document}